\documentclass[prl,aps,twocolumn,superscriptaddress]{revtex4}

\usepackage{dcolumn}
\usepackage{bm}
\usepackage{amssymb}
\usepackage{color}
\usepackage[pdftex]{graphicx}

\usepackage{amsmath}
\usepackage[linesnumbered,ruled]{algorithm2e}

\usepackage{url}
\usepackage[section]{placeins}
\usepackage[colorlinks=true,linkcolor=blue,citecolor=blue]{hyperref}

\begin{document}


\title{Nonequilibrium Dynamics of Gating-Induced Resistance Transition \\ in Charge Density Wave Insulators}

\author{Sheng Zhang}
\affiliation{Department of Physics, University of Virginia, Charlottesville, VA 22904, USA}

\author{Yunhao Fan}
\affiliation{Department of Physics, University of Virginia, Charlottesville, VA 22904, USA}

\author{Gia-Wei Chern}
\affiliation{Department of Physics, University of Virginia, Charlottesville, VA 22904, USA}

\date{\today}

\begin{abstract}
We present a comprehensive numerical investigation of the gate-induced insulator-to-metal transition in the charge-density-wave (CDW) phase of the Holstein model. Large-scale Brownian-dynamics simulations are performed, in which the forces acting on the lattice degrees of freedom are evaluated using the nonequilibrium Green’s function formalism. We demonstrate that the onset of CDW instability requires a threshold bias voltage set by the energy of in-gap edge modes. At sufficiently large voltages, the system undergoes an abrupt transition to a metallic state, reminiscent of dielectric breakdown. In the intermediate-voltage regime, our simulations reveal that the transition to a low-resistance state is initiated by the nucleation of a thin conducting layer at the gated electrode. The resulting metal–insulator interface subsequently propagates across the system under the applied bias, leading to the growth of a metallic domain. We further analyze the voltage- and temperature-dependent dynamics of the associated domain walls.
\end{abstract}

\maketitle

Complex adaptive materials exhibiting multiple resistive states are widely explored as functional building blocks for next-generation electronic and information technologies. The operation of conventional metal-oxide semiconductor field-effect transistors (MOSFETs) relies on the electrostatic charging of free carriers in thermal equilibrium, which fundamentally limits the subthreshold swing to 60 mV per decade, known as the Boltzmann limit. In contrast, electrically controllable switching between distinct many-body states with different resistances offers a route toward high-speed and low-power transistor operation. In particular, it has been proposed that ultrafast switching can be realized through the collective electronic response associated with an insulator-to-metal transition (IMT)~\cite{mott90,imada98,dobrosavljevic12}. Along these lines, the emerging field of “Mottronics” seeks to implement electronic and logic functionalities by harnessing strong electron correlations in Mott insulators~\cite{nakano12,shukla15,song16}. In these materials, the high-resistance insulating state originates from electron localization driven by short-range Coulomb repulsion, while a transition to a low-resistance metallic state can be induced via a first-order phase transition by reducing the effective electron–electron interaction through electrostatic doping.

Charge-density-wave (CDW) systems~\cite{gruner88,thorne96,balandin21} represent another class of correlated many-body states that has attracted substantial interest owing to their rich physical properties and potential electronic and optoelectronic applications. A CDW phase is a macroscopic quantum state characterized by a periodic modulation of the electronic charge density accompanied by a commensurate lattice distortion. Early studies of CDW materials focused primarily on the sliding dynamics of quasi-one-dimensional incommensurate states, which exhibit nonlinear conductivity under low applied electric fields~\cite{thorne96}. Subsequent work has uncovered a variety of additional phenomena, including giant dielectric responses, multistable conductive states, and proximity to unconventional superconductivity. Interest in CDW physics has been revitalized in recent years by the discovery of quasi-two-dimensional van der Waals materials, in which CDW order can persist up to and beyond room temperature~\cite{hellmann12,porer14,stojchevska14,yu15,samnakay15,xi15,vaskivskyi15,vogelgesang18,cho16}. Owing to their sensitivity to external stimuli such as temperature, strain, and applied bias, these quasi-two-dimensional CDW materials have emerged as a promising platform for multifunctional device applications.

In particular, electrically driven transitions between distinct CDW phases, as well as CDW-to-metal transitions, have been experimentally demonstrated in the 1T polymorph of tantalum disulfide (TaS$_2$)~\cite{hollander15,vaskivskyi16,grisafe18,li16,geremew19,zheng17}. These transitions occur on nanosecond timescales, rendering this material a compelling candidate for high-speed, energy-efficient electronic devices~\cite{zheng17,liu16,khitun18}. The voltage- or electric-field-induced IMT is intrinsically a highly complex nonequilibrium process involving the interplay of carrier dynamics, lattice distortions, thermal transport, and electron correlations. The electrically driven CDW-to-metal transition is no exception. Recent experiments suggest, however, that this process is primarily governed by the nucleation and propagation of complex domain walls~\cite{vaskivskyi16}, distinguishing it qualitatively from resistive switching in complex oxides or chalcogenide glasses, where the transition proceeds through highly inhomogeneous intermediate states~\cite{waser07,sawa08,waser09,kim11,jeong12,lee15,janod15}. 

Despite its fundamental significance and technological relevance, theoretical modeling of electrically induced CDW transitions has largely remained at a phenomenological level. This challenge arises from the intrinsically multiscale nature of the problem. On the microscopic level, one must describe the nonequilibrium electronic response and its feedback on the lattice degrees of freedom, while on mesoscopic and macroscopic scales, large-scale real-space simulations are required to capture transient pattern formation and domain-wall nucleation and propagation. For systems with strong electronic correlations, the need for appropriate many-body techniques further exacerbates the computational complexity.

In this work, we present large-scale dynamical simulations of the CDW phase transition, explicitly resolving the interplay between lattice dynamics and nonequilibrium electronic effects. We consider the adiabatic limit of the Holstein model, in which the lattice degrees of freedom are treated as classical variables, allowing the electronic subsystem to be solved exactly at each time step. By performing extensive quantum Brownian-dynamics simulations with forces evaluated using the nonequilibrium Green’s function formalism, we demonstrate a reversible, gate-induced CDW-to-metal transition driven by domain-wall motion. Furthermore, we construct a dynamical phase diagram and systematically characterize the voltage and temperature dependence of the domain-wall dynamics. 

\begin{figure}[t]
\includegraphics[width=0.99\columnwidth]{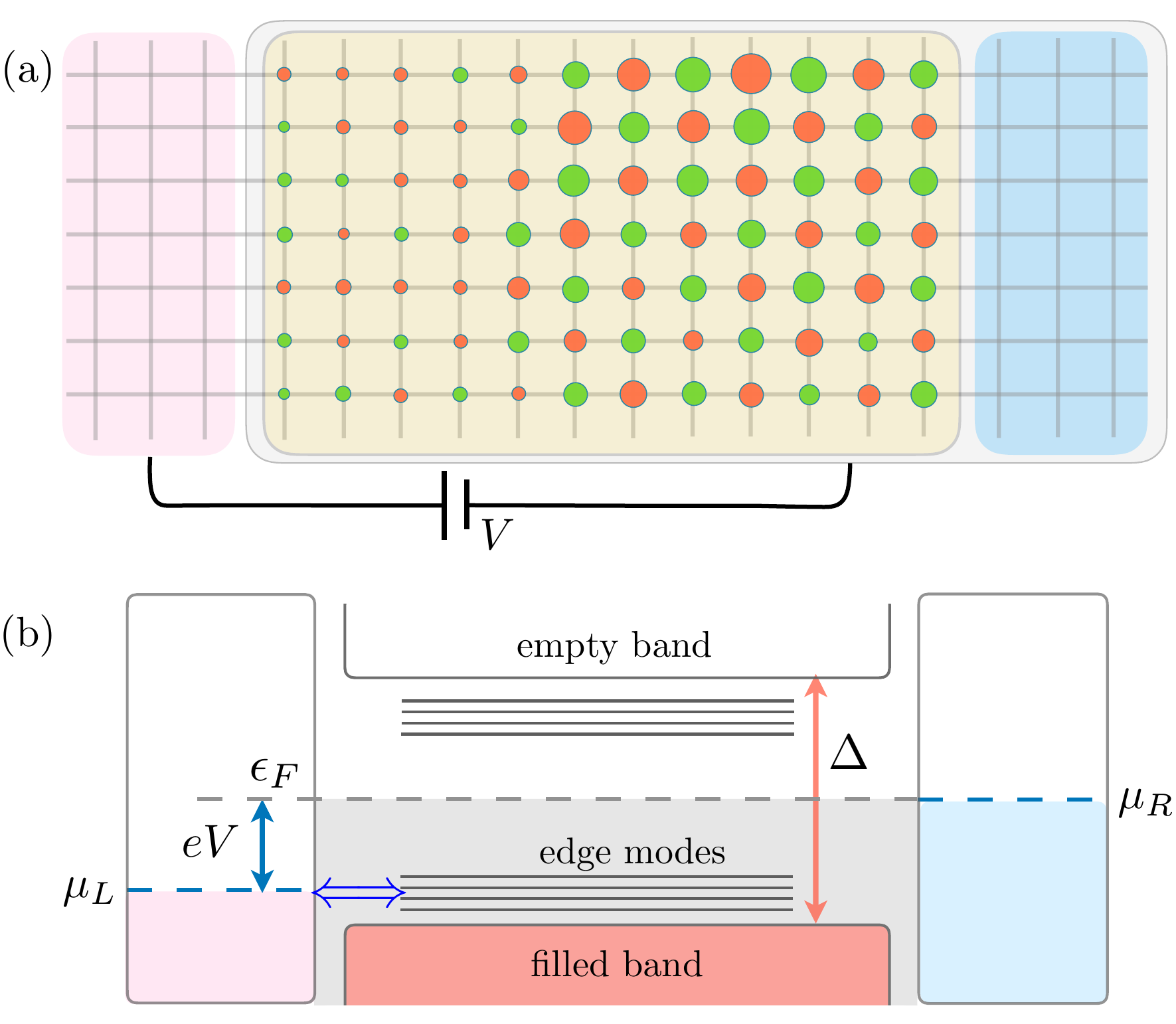}
\caption{(Color online)  
\label{fig:schematic} (a) Schematic illustration of the gate-induced transition from a charge-density-wave (CDW) insulator to a metallic state in the adiabatic Holstein model.  (b) Energy-level alignment of the two electrodes and the central Holstein system. In the CDW phase, the system is a band insulator with an energy gap $E_g = 2 g Q_0$. The Fermi level $\epsilon_F$ in the bulk and the chemical potential $\mu_R$ of the right electrode are aligned at the center of the gap. The chemical potential of the left electrode is lowered relative to $\epsilon_F$ by the applied gate voltage, $\mu_L = \epsilon_F - eV$. 
}
\end{figure}

We consider a capacitor geometry in which a square-lattice Holstein model is sandwiched between two electrodes, as illustrated in Fig.~\ref{fig:schematic}(a). The right electrode serves as a substrate, while a gate voltage $V$ is applied to the left electrode. The total Hamiltonian is given by $\mathcal{H} = \mathcal{H}_{\rm Hols} + \mathcal{H}_{\rm res}$, where $\mathcal{H}_{\rm Hols}$ describes the square-lattice Holstein model in the central region and  $\mathcal{H}_{\rm res}$ accounts for the electrode and reservoir degrees of freedom. The Holstein Hamiltonian reads~\cite{holstein59}
\begin{eqnarray}
	\label{eq:Holstein}
	\hat{\mathcal{H}}_{\rm Hols}=-t_{\rm nn} \sum_{\langle i j \rangle} \hat{c}_{i}^{\dagger} \hat{c}_{j} - g\sum_{i} Q_{i} \left( \hat{n}_{i} -\frac{1}{2} \right) 
\end{eqnarray}
where $\hat{c}_{i}/\hat{c}^\dagger_i$ is the annihilation/creation operators of spin-less electron at site-$i$, and  $\hat{n}_{i} \equiv \hat{c}^\dagger_{i} \hat{c}^{\;}_{i}$ is the corresponding number operator, $Q_i$ describes a local structural distortion at site-$i$, such as the breathing mode of the oxygen octahedron. The first term describes nearest-neighbor hopping $t_{\rm nn}$ of electrons, and the second term denotes phonon-electron interaction with a coupling constant $g$. As discussed above, here we treat the Holstein phonon $Q_i$ as classical variables with the following elastic energy~\cite{holstein59}
\begin{eqnarray}
	\mathcal{V} = \frac{k}{2} \sum_{i}  Q_{i}^2 + \sum_{\langle ij \rangle} \kappa \, Q_i Q_j \, ,  
\end{eqnarray}
where $k$ is the effective on-site spring constant, and $\kappa$ describes a nearest-neighbor repulsion between the local distortions. The Holstein model is a paradigmatic electron–phonon system and has been extensively employed to study polarons, superconductivity, and CDW physics~\cite{bonca99,golez12,mishchenko15,zhang19,costa18,marsiglio90,vekic92,zheng97,cohen20}. 
In the adiabatic limit, analogous to the Born–Oppenheimer approximation commonly used in {\em ab initio} molecular dynamics~\cite{marx09}, the electronic degrees of freedom are assumed to relax much faster than the lattice dynamics. We therefore model the time evolution of the lattice distortions using over-damped Langevin, or the Brownian dynamics (BD)~\cite{cohen20,goetz21,michielsen97}
\begin{eqnarray}
	\frac{d Q_i}{dt} = -\frac{1}{\gamma} \left( -\frac{\partial \mathcal{V}}{\partial Q_i} +  F^e_i \right) + \zeta_i(t),
\end{eqnarray}
where $\gamma$ is an effective friction constant, and $\eta_i(t)$ denotes a stochastic force described by a delta-correlated stationary Gaussian process. The driving forces on the right-hand side consist of two contributions: the first term $-\partial \mathcal{V} / \partial Q_i$ accounts for the classical elastic restoring force, while $F^e_i$ represents the forces due to electrons. The driving force of a conservative system is given by the derivative of potential energy: $F_i = - \partial E / \partial Q_i$, which is indeed the case for the classical potential $\mathcal{V}(\{Q_i\})$.  For electron forces in thermal equilibrium, the effective potential energy due to electrons is given by $E_e = \langle \hat{\mathcal{H}}_{\rm Hols} \rangle = {\rm Tr}\bigl( \hat{\rho}_{\rm eq} \hat{\mathcal{H}}_{\rm Hols} \bigr) $. In contrast, for a driven nonequilibrium electronic system, the energy is not well defined and the resulting forces are nonconservative. Nevertheless, the electronic force can still be evaluated using a generalized Hellmann–Feynman theorem~\cite{diventra00,todorov01,lu12,todorov10,dundas09}
\begin{eqnarray}
	\label{eq:force}
	F^e_i = -\biggl\langle \frac{\partial \hat{\mathcal{H}}_{\rm Hols}}{\partial Q_i} \biggr\rangle = g \langle \hat{n}_i \rangle 
\end{eqnarray}
where the expectation value $\langle \cdots \rangle$ is evaluated with respect to the quasi-steady-state, yet out-of-equilibrium, electron subsystem.

\begin{figure*}[t]
\includegraphics[width=1.99\columnwidth]{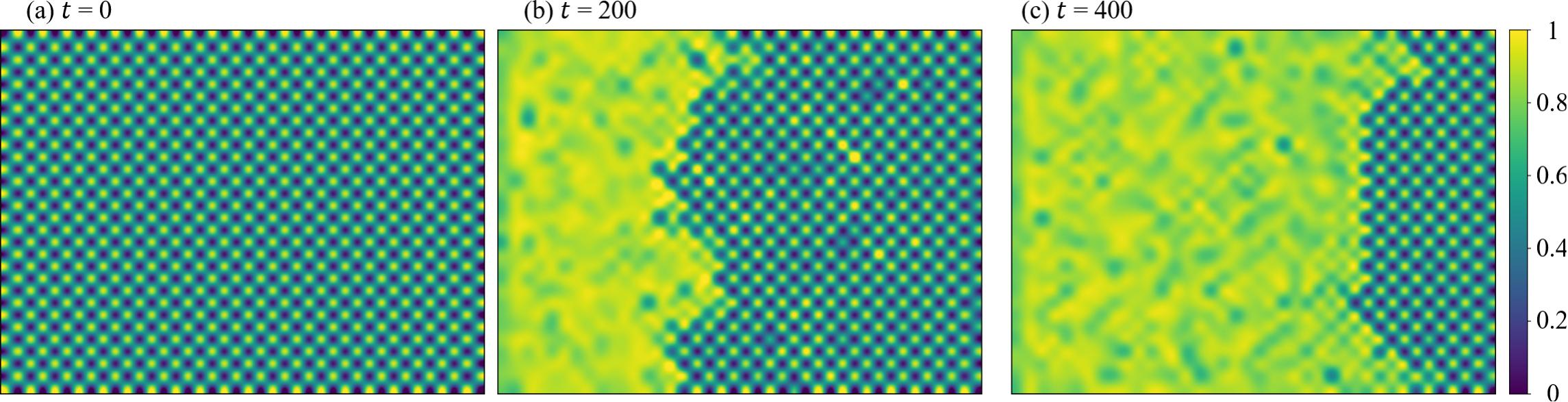}
\caption{(Color online)  
\label{fig:snapshot} NEGF-BD simulations of a driven Holstein model on a $40 \times 30$ square lattice. The bias voltage is applied along the longitudinal $x$ direction. Panels (a)--(c) display snapshots of the local electronic density, $n_i = \langle \hat{c}^\dagger_i \hat{c}^{\,}_i \rangle$, at different simulation times. The parameters used are: electron-phonon coupling constant $g=1.5 t_{\rm nn}$, the effective spring constant $k_1=0.67g$, nearest-neighbor elastic coupling $\kappa=0.12 g$, damping constant $\lambda=0.2g$, $\Gamma_{\text{lead}}=1.0$, $\Gamma_{\text{bath}}=0.001$, $k_BT=0.1$, and the bias voltage $eV = 2.5 t_{\rm nn}$.
}
\end{figure*}

We next employ the nonequilibrium Green's function (NEGF) method~\cite{meir92,jauho94,haug08,diventra08} to evaluate the electronic expectation values entering the lattice equations of motion. The electrodes and reservoir degrees of freedom are modeled explicitly by the Hamiltonian
\begin{eqnarray}
	\mathcal{H}_{\rm res} = \sum_{\xi, i} \varepsilon_\xi \, d^\dagger_{i, \xi} d^{\;}_{i, \xi} - \sum_{i, \xi} V_{\xi, i} \bigl(d^\dagger_{i, \xi} c^{\,}_{i} + {\rm h.c.} \bigr). \quad
\end{eqnarray} 
Here $d_{i, \xi}$ represents non-interacting fermions from the bath (for $i$ inside the bulk) or the leads (for $i$ on the two open boundaries), and $\xi$ is a continuous reservoir quantum number, such as the band index of the leads.

After integrating out the reservoir fermions in both leads and bath, the retarded Green's function matrix for the central region is given by $\mathbf G^r(\epsilon) = (\epsilon \mathbf I - \mathbf H - \bm \Sigma^r)^{-1}$, where $H_{ij} =  t_{ij}  - g \delta_{ij} Q_i$ is the tight-binding matrix of the Holstein model~Eq.~(\ref{eq:Holstein}) and 
\begin{eqnarray}
	\Sigma^r_{i j}(\epsilon) = \delta_{ij} \sum_\xi  |V_{i,\xi} |^2/ ( \epsilon - \epsilon_\xi + i 0^+)
\end{eqnarray}
is the dissipation-induced self-energy. The resultant level-broadening matrix given by $\bm\Gamma = i (\bm\Sigma^r - \bm\Sigma^a)$ is diagonal with $\Gamma_{ii} = \pi  \sum_\xi  |V_{i, \xi}|^2 \delta(\epsilon - \epsilon_\xi)$. For simplicity, we adopt the wide-band limit, assuming a flat reservoir density of states, which yields energy-independent broadening parameters. 
Next, using the Keldysh formula, the lesser Green's function is obtained from the retarded/advanced Green's functions: $\mathbf G^{<}(\epsilon) = \mathbf G^r(\epsilon) \bm\Sigma^{<}(\epsilon) \mathbf G^a(\epsilon)$, and the lesser self-energy is related to the $\Sigma^{r/a}$ through dissipation-fluctuation theorem: $\Sigma^{<}_{i  j}(\epsilon) = 2 i \,\delta_{ij} \, \Gamma_{i} \, f_{\rm FD}(\epsilon - \mu_i)$. Here $\Gamma_i = \Gamma_{\rm lead}$ or $\Gamma_{\rm bath}$ depending on whether site-$i$ is at the boundaries or in the bulk. The local chemical potential $\mu_i = \epsilon_F$ for the bath, and $\mu_i = \mu_{L/R} $ for the two electrodes.  Finally, the nonequilibrium force in Eq.~(\ref{eq:force}) is proportional to the on-site electron density $\langle \hat{n}_i \rangle =  \langle c^\dagger_i(t) c^{\,}_j(t) \rangle$, which is the diagonal element of the equal-time lesser Green's function
\begin{eqnarray}
	F^e_i(t)  = g \, G^<_{ii}(t, t) = g \int_{-\infty}^{+\infty} G^<_{ii}(\epsilon; t) \,d\epsilon.
\end{eqnarray}
In practice, the energy integral is evaluated using a Riemann sum with step size $\Delta \epsilon = 0.003$, corresponding to up to $4\times10^{3}$ energy grid points.

We apply the NEGF-BD method outlined above to simulate the gating-induced IMT in the Holstein model on a $40\times 30$ square lattice. The initial state is prepared via equilibrium Brownian-dynamics simulations at half filling, $\overline{n} \equiv  \sum_{i} \hat{n}_{i} / N = 0.5$, yielding an initial state with CDW order on the lattice; see Fig.~\ref{fig:snapshot}(a). In the presence of perfect CDW order, the electronic band structure is given by $E_{\pm}(\mathbf k) = \pm \sqrt{\epsilon_{\mathbf k}^2 +g^2\, Q_0^2}$, where $\epsilon_{\mathbf k}$ is the energy dispersion of the square-lattice tight-binding Hamiltonian, and $Q_0$ is the amplitude of the lattice distortion in the CDW state. Importantly, an energy gap $\Delta = 2 g \, Q_0$ opens at the Fermi level, and at half filling the lower band $E_-(\mathbf k)$ is occupied, rendering the system a band insulator.

We then apply a voltage bias $V$. As illustrated in Fig.~\ref{fig:schematic}(b), the chemical potential of the substrate (right electrode) and the central region are fixed at the zero $\epsilon_F = \mu_R =0$, which lies at the center of the CDW gap, while the chemical potential of the left electrode is shifted to $\mu_L = - eV$. 
Fig.~\ref{fig:snapshot} illustrates a representative phase transformation induced by a gate voltage $eV = 2.5$, where the color scale indicates the local electron density $ \langle \hat{n}_i \rangle$. Other simulation parameters are:  electron-phonon coupling constant $g=1.5 t_{\rm nn}$, on-site spring constant $k_1=0.67g$, nearest-neighbor coefficient $\kappa=0.12 g$, damping constant $\lambda=0.2g$, $\Gamma_{\text{lead}}=1.0$, $\Gamma_{\text{bath}}=0.001$, and temperature $k_BT=0.1$.

\begin{figure}[t]
\includegraphics[width=0.85\columnwidth]{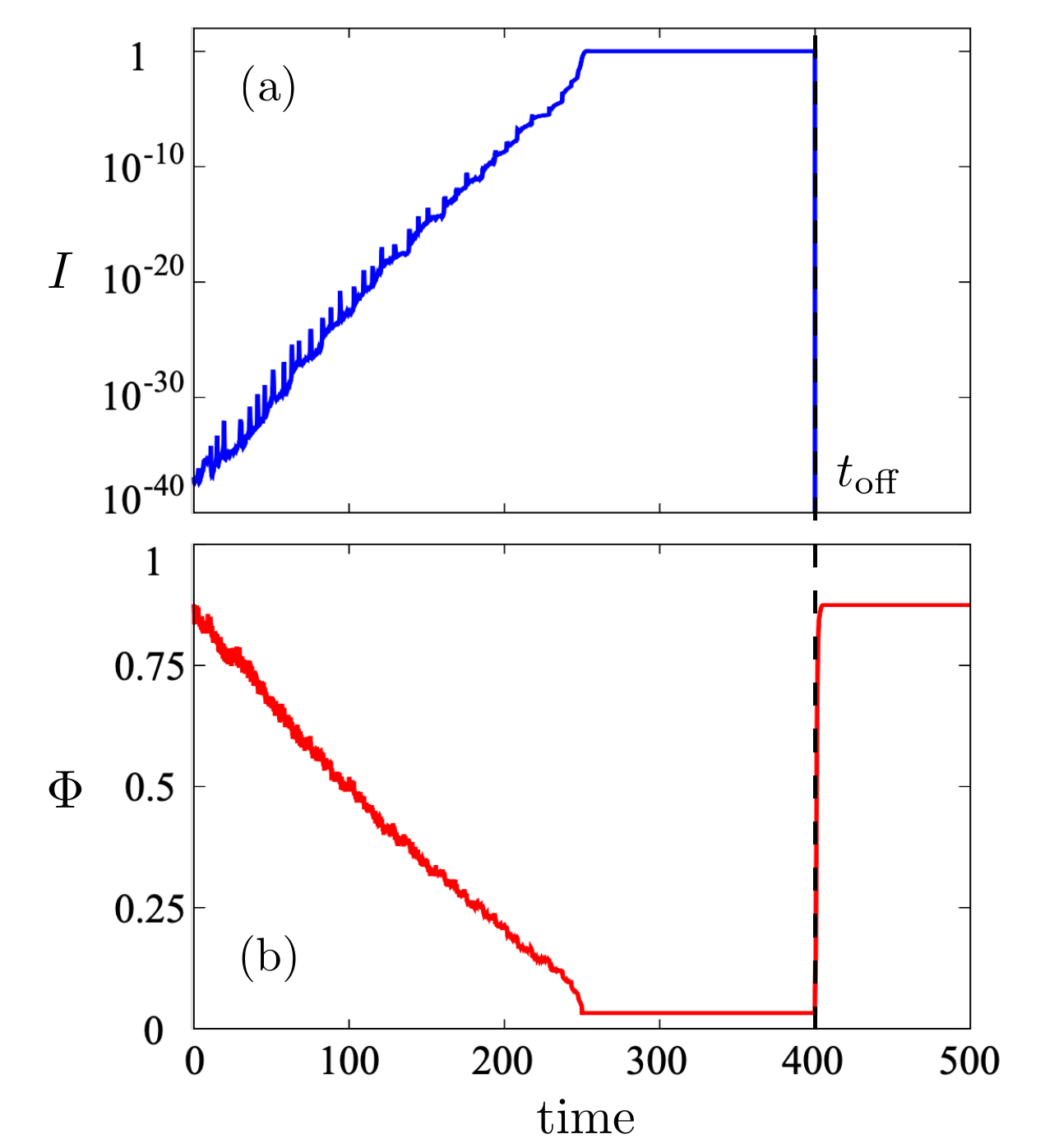}
\caption{(Color online)  
\label{fig:cdw-time} Time dependence of (a) the transmission current $I$ and (b) the charge-density-wave (CDW) order parameter $\Phi$ at zero temperature. A constant voltage $eV = 2.5 t_{\rm nn}$ is applied to the Holstein system over the time interval $0\leq t \leq t_{\text{off}}=400$. 
}
\end{figure}

As shown in Fig.~\ref{fig:snapshot}, the applied gate voltage destabilizes the CDW state and induces the formation of a metallic domain as electrons are extracted through the gate electrode with a lower chemical potential. The resulting CDW–metal interface is subsequently driven toward the substrate electrode. To further characterize the nonequilibrium phase transformation, we compute the transmission current of the driven electronic system using the NEGF formalism~\cite{meir92,jauho94,haug08,diventra08} 
\begin{eqnarray}
	I = \int  {\rm Tr}({\bm \Gamma}_R\, {\bf G}^r \, {\bm \Gamma}_L \, {\bf G}^{a} ) [f_L(\epsilon) - f_R(\epsilon)] d\epsilon, 
\end{eqnarray}
where $\bm\Gamma_{L, R}$ are the diagonal broadening matrices, and $f_{L, R}(\epsilon) = f_{\rm FD}(\epsilon - \mu_{L, R})$ are the Fermi-Dirac functions. Fig.~\ref{fig:cdw-time}(a) displays the time dependence of the transmission current $I$ under a bias voltage $eV=2.5$ at $T \to 0$. The approximately linear behavior in the semilogarithmic plot indicates an exponential growth of the current, $I \sim \exp(c\, t)$, with $c$ a numerical constant. At such low temperatures, the phase transformation proceeds essentially as a downhill process, with excess energy dissipated through both the lattice damping $\gamma$ and the electronic reservoirs. Notably, as the CDW–metal interface advances across the lattice one layer at a time, this discrete motion gives rise to small oscillatory features superimposed on the exponential current growth.
Fig.~\ref{fig:cdw-time}(b) shows the temporal evolution of the CDW order parameter, 
\begin{eqnarray}
	\Phi = \sum_i \langle \hat{n}_i \rangle \exp(i \mathbf Q \cdot \mathbf r_i),
\end{eqnarray}
where $\mathbf Q_i = (\pi/a, \pi/a)$ is the wave vector characterizing the checkerboard CDW pattern on the square lattice. The nearly linear decay of $\Phi$ during the transformation is consistent with a roughly constant velocity of the propagating CDW–metal domain wall. 

Our simulations further reveal that the metallic state is stable only in the presence of the bias voltage, indicating that the insulator–metal transition is reversible. As shown in Fig.~\ref{fig:cdw-time}, the CDW order parameter recovers its initial value once the gate voltage is switched off. In contrast to the voltage-driven transition, which proceeds through intermediate states with a propagating domain wall, the recovery to the CDW state occurs almost instantaneously, signaling a bulk instability. It is worth noting that the complete restoration of the CDW order to its maximal value is a finite-size effect; for larger systems, the reverse transition is expected to generate multiple CDW domains with opposite phases.

\begin{figure}[t]
\includegraphics[width=0.9\columnwidth]{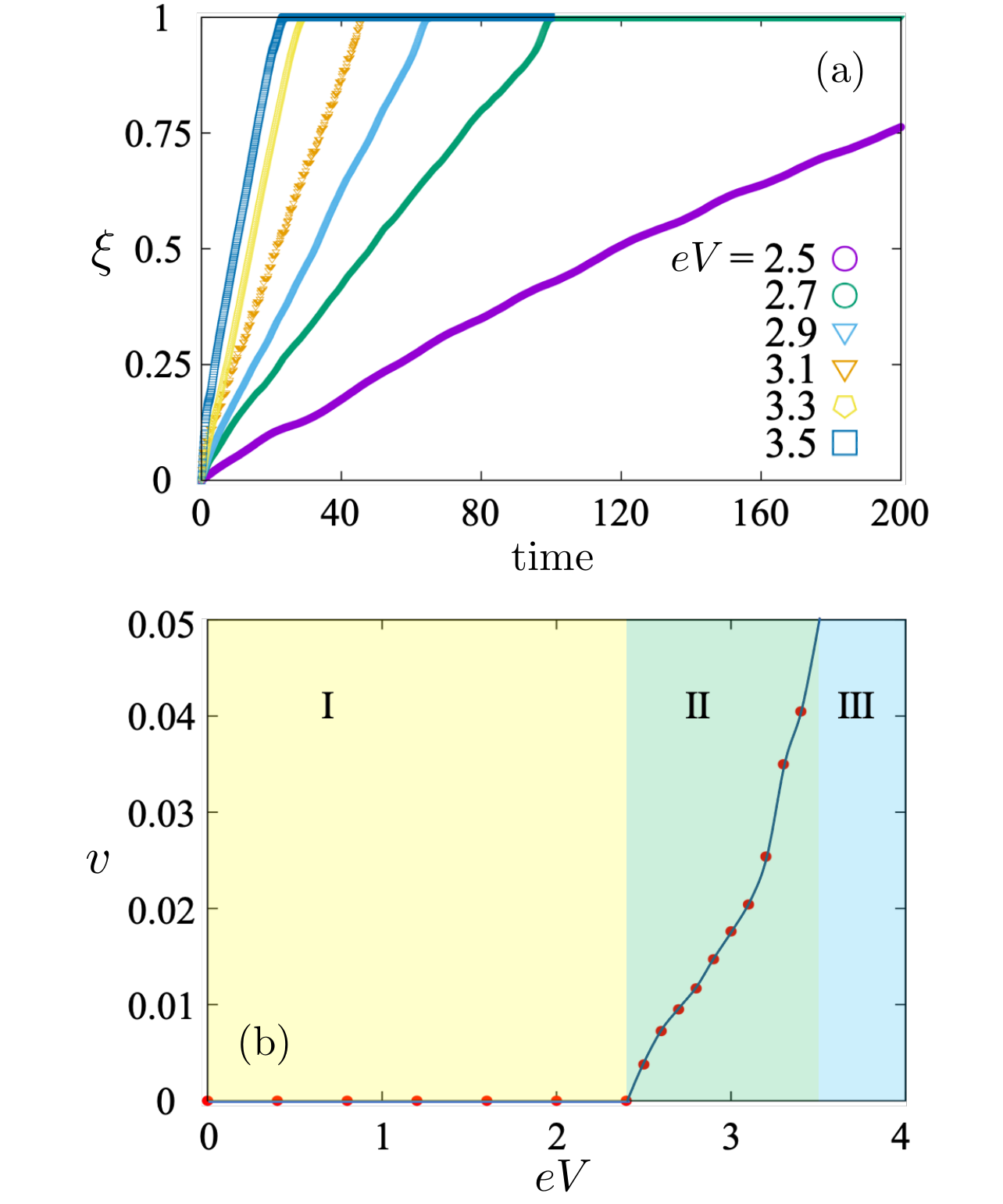}
\caption{(Color online)  
\label{fig:velocity-eV} (a) Average position $\xi$ of the CDW-metal domain wall as a function of time for different gating voltages at zero temperature. (b)Velocity $v$ of the CDW–metal domain wall as a function of the applied gate voltage~$eV$, extracted from the average slope of the $\xi(t)$ curves shown in panel~(a).
}
\end{figure}

We next examine the dependence of the phase transformation dynamics on the applied gate voltage. Fig.~\ref{fig:velocity-eV}(a) shows the average position $\xi$ of the CDW-metal interface as a function of time at different bias voltages. As expected, increasing the gate voltage accelerates the domain-wall motion. The corresponding average velocity, extracted from a linear fit to $\xi(t)$, is plotted in Fig.~\ref{fig:velocity-eV}(b) as a function of the bias voltage. Importantly, one can identify three different dynamical regimes from this result. For gate voltages below a threshold value $eV_{\rm th} \sim 2.4$, the CDW state remains stable. In regime-II, the applied bias induces the formation of a CDW–metal interface whose propagation velocity increases monotonically with voltage. Finally, above a second critical voltage, the system undergoes an essentially instantaneous transition to the metallic phase, similar to a dielectric breakdown. 

This dynamical phase diagram can be understood from the energy-level alignment shown in Fig.~\ref{fig:schematic}(b), which also provides an explanation of the phase transition mechanisms. At low bias, the chemical potential of the gate electrode $\mu_L$ lies within the CDW gap and is therefore ineffective in driving a phase transition. As the voltage increases such that $\mu_L$ overlaps with the energy levels of the in-gap states that are localized at the left edge, the resonant coupling leads to an instability of the CDW. As electrons are drained from the gate due to a smaller chemical potential, a metallic layer is nucleated near the left edge. The subsequent phase transformation proceeds through the expansion of this metallic domain, corresponding to the dynamical regime-II in Fig.~\ref{fig:velocity-eV}(b). At still higher voltages, when $\mu_L$ drops below the valence-band edge, the gate electrode couples directly to bulk electronic states, enabling efficient charge transport and triggering an abrupt transition to the metallic phase.

\begin{figure}[t]
\includegraphics[width=0.9\columnwidth]{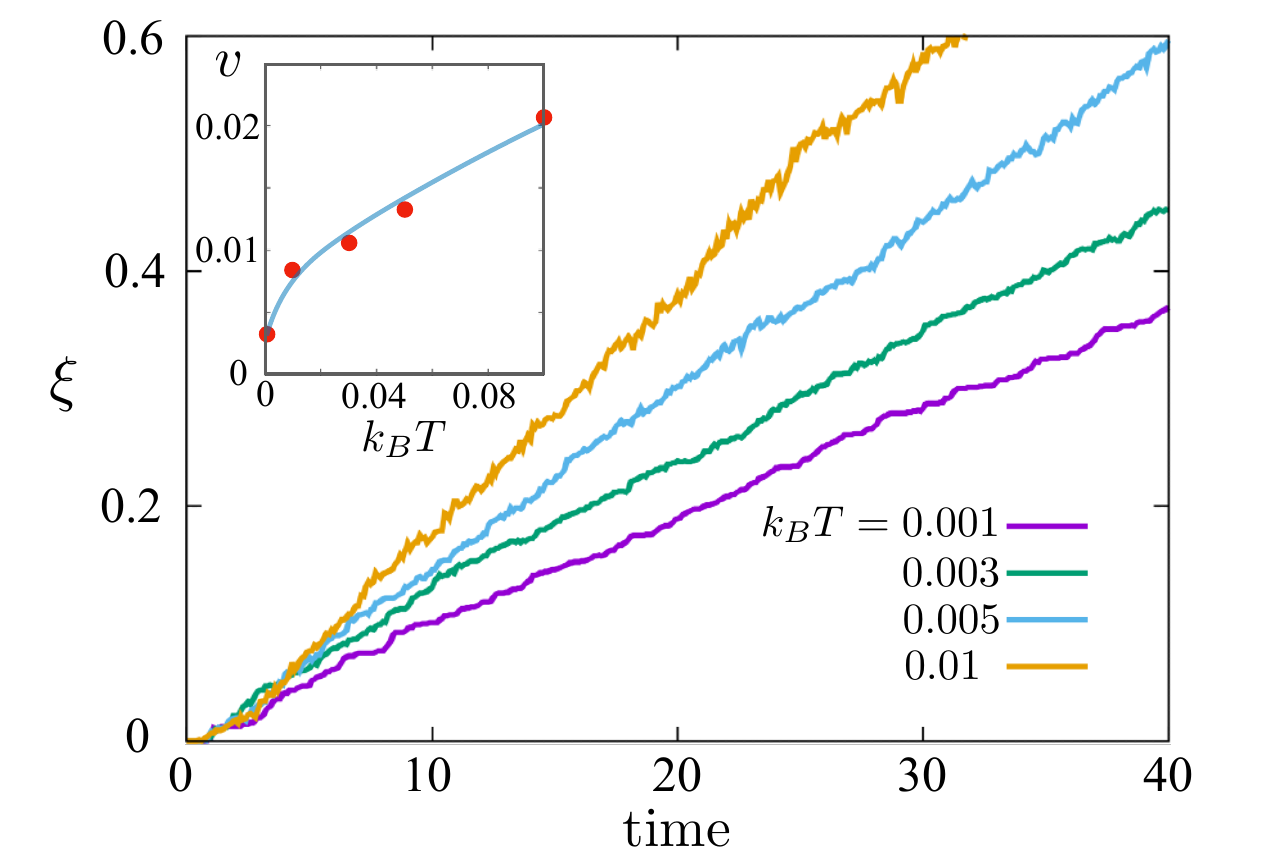}
\caption{(Color online)  
\label{fig:velocity-kT} Average position $\xi$ of the CDW-metal interface as a function of time for several temperatures under a constant driving voltage $eV=2.5 t_{\rm nn}$. The inset shows the CDW-metal domain-wall velocity as a function of temperature. 
}
\end{figure}


The domain wall propagation in the $T \to 0$ limit discussed above is governed by a relaxation dynamics dominated by energy dissipation. It is instructive to note that the effective potential for the local lattice distortion resembles a double-well structure: one minimum at $Q_i = \pm Q_0$ (depending on the sublattice) and the other one at $Q_i \sim 0$ corresponding to the metallic phase. The actual energy landscape of the full system, however, is considerably more intricate than this simple picture. The layer-by-layer progression of the metallic domain suggests that each domain-wall configuration constitutes a quasi-stable state in a high-dimensional configuration space. As energy is continuously drained through the gate, the decay of a local minimum triggers the advancement of the domain wall by one layer. In the presence of thermal fluctuations, the system can transition between local minima via thermally assisted tunneling rather than waiting for the decay of the current metastable state. Consequently, one expects an enhancement of the domain-wall mobility at finite temperature. This expectation is borne out in our simulations, as shown in Fig.~\ref{fig:velocity-kT}. With the exception of the lowest-temperature point, the domain-wall velocity exhibits an approximately linear increase with temperature. 

In conclusion, we have performed large-scale NEGF–Brownian dynamics simulations to investigate the gate-induced CDW–metal transition in a prototypical electron–phonon model. Beyond the well-known electrical breakdown at large bias, we uncover an intermediate dynamical regime characterized by metastable configurations in which metal–insulator domain walls propagate through the system. The initial instability of the CDW is triggered by the nucleation of a metallic layer at the gated interface, after which the metal–CDW boundary advances with an approximately constant velocity that increases with both voltage bias and temperature. Notably, the transition is reversible: once the gate voltage is removed, the system rapidly relaxes back to the CDW phase. 

Our work focuses on the interplay between atomic lattice dynamics and the out-of-equilibrium electrons, and it highlights the mechanisms underlying the dynamical instability of CDW states under gating. Nevertheless, CDW phase transitions in real materials—particularly quasi-2D van der Waals compounds—are considerably more complex than what is captured by the adiabatic Holstein model employed here. In particular, the present simulations neglect electron–electron correlations, which are believed to play an essential role in materials such as 1T-TaS$_2$ that are intensively studied experimentally. More sophisticated many-body techniques, such as Hartree-Fock or Gutzwiller methods, are required to properly model the collective electron behaviors in CDW described by, e.g. the Holstein-Hubbard model~\cite{zhong92,weber18}. For instance, self-consistent Hartree–Fock calculations combined with NEGF have been used to study voltage-driven spin-density–wave patterns in Hubbard systems~\cite{ribeiro16,li17,dutta20}. A full dynamical treatment of CDW transitions, however, would require integrating these techniques with real-space lattice dynamics, a task that remains computationally challenging. Machine-learning–based surrogate models may provide a promising route toward such multiscale simulations of complex CDW dynamics.

\bigskip

\begin{acknowledgments}
The author thanks G. Kotliar, and Jong~Han for fruitful discussions. This work is supported by the US Department of Energy Basic Energy Sciences under Award No. DE-SC0020330. The author also acknowledge the support of Advanced Research Computing Services at the University of Virginia.
\end{acknowledgments}

\newpage

\end{document}